\documentclass[draftcls,12pt,onecolumn,oneside]{IEEEtran}

\usepackage{amsmath,amssymb} \interdisplaylinepenalty=2500
\usepackage{color}
\usepackage{subfigure}
\usepackage{epsfig}
\usepackage{graphicx}
\usepackage{dsfont}
\usepackage{verbatim}
\usepackage{amsthm}
\usepackage{amsmath}
\usepackage{setspace}
\usepackage{mathtools}
\usepackage{enumerate}

\usepackage{tabulary}
\usepackage{booktabs}
\usepackage{bbm}

\usepackage[table]{xcolor}
\usepackage{color, colortbl}
\usepackage{multirow,bigdelim}
\usepackage{amstext}
\usepackage{array}
\usepackage{cite}

\newcount\rowc

\makeatletter
\def\ttabular{%
\hbox\bgroup
\let\\\cr
\def\rulea{\ifnum\rowc=\@ne \hrule height 1.3pt \fi}
\def\ruleb{
\ifnum\rowc=1\hrule height 1.3pt \else
\ifnum\rowc=6\hrule height \heavyrulewidth 
   \else \hrule height \lightrulewidth\fi\fi}
\valign\bgroup
\global\rowc\@ne
\rulea
\hbox to 10em{\strut \hfill##\hfill}%
\ruleb
&&%
\global\advance\rowc\@ne
\hbox to 10em{\strut\hfill##\hfill}%
\ruleb
\cr}
\def\endttabular{%
\crcr\egroup\egroup}

\newtheorem{definition}{Definition}
\newtheorem{theorem}{Theorem}
\newtheorem{lemma}[theorem]{Lemma}

\newenvironment{remark}{\textit{Remark: }}{}

\newif\ifcomments
\commentstrue 

\newcommand{\calM}{\mathcal{M}}

\newcommand{\calT}{\mathcal{T}}
\newcommand{\calU}{\mathcal{U}}

\newcommand{\bfA}{\mathbf{A}}

\newcommand{\bfD}{\mathbf{D}}
\newcommand{\bfE}{\mathbf{E}}

\newcommand{\bfG}{\mathbf{G}}

\newcommand{\bfU}{\mathbf{U}}

\newcommand{\bfW}{\mathbf{W}}

\DeclarePairedDelimiterX{\norm}[1]{\lVert}{\rVert}{#1}

\newcommand{\Mod}[1]{\ (\text{mod}\ #1)}

\newcommand{\nFile}{K}
\newcommand{\sFile}{L}
\newcommand{\nNode}{N}
\newcommand{\nSysNode}{M}
\newcommand{\queries}{\mathcal{Q}}

\newcommand{\answer}{A}
\newcommand{\secrecy}{S}

\newcommand{\compkprime}{\bar{k'}}
\newcommand{\compk}{\bar{k}}

\newcommand{\Fq}{\mathds{F}_{q}}

\newcommand{\nodeset}{\mathcal{N}}

\newcommand{\findex}{\kappa}

\newcolumntype{L}{>{$}l<{$}}

\definecolor{Gray}{gray}{0.9}
\definecolor{LightCyan}{rgb}{0.88,1,1}
\definecolor{HoneydewTwo}{rgb}{0.96,0.92,0.88}

\begin{document}

\title{Linear Symmetric Private Information Retrieval for MDS Coded Distributed Storage with Colluding Servers}
\author{\IEEEauthorblockN{Qiwen~Wang, and~Mikael~Skoglund} \\
    \IEEEauthorblockA{School of Electrical Engineering, KTH Royal Institute of Technology} \vspace{-0.2em}
    \\Email: \{qiwenw, skoglund\}@kth.se \vspace{-0.3em}}

\maketitle

\thispagestyle{plain}
\pagestyle{plain}

\vspace{-1cm}
\begin{abstract}
The problem of \emph{symmetric private information retrieval (SPIR)} from a coded database which is distributively stored among colluding servers is studied. Specifically, the database comprises $K$ files, which are stored among $N$ servers using an $(N,M)$-MDS storage code. A user wants to retrieve one file from the database by communicating with the $N$ servers, without revealing the identity of the desired file to any server. Furthermore, the user shall learn nothing about the other $K-1$ files in the database. In the $T$-colluding SPIR problem (hence called TSPIR), any $T$ out of $N$ servers may collude, that is, they may communicate their interactions with the user to guess the identity of the requested file.
We show that for linear schemes, the information-theoretic capacity of the MDS-TSPIR problem, defined as the maximum number of information bits of the desired file retrieved per downloaded bit, equals $1-\frac{M+T-1}{N}$, if the servers share common randomness (unavailable at the user) with amount at least $\frac{M+T-1}{N-M-T+1}$ times the file size. Otherwise, the capacity equals zero. We conjecture that our capacity holds also for general MDS-TSPIR schemes.
\end{abstract}

\section{Introduction}
The situation where a user wants to retrieve a file from a database without revealing the identity of the requested file is known as the problem of private information retrieval (PIR). It is shown that if the database is stored at a single server, the only possible scheme for the user is to download the entire database to guarantee information-theoretic privacy \cite{chor1995private,chor1998private}, which is inefficient in practice. It is further shown that the communication cost can be reduced in sublinear scale by replicating the database at multiple non-colluding servers \cite{chor1998private}.
To further protect the privacy of the database, symmetric private information retrieval (SPIR) is introduced~\cite{gertner1998protecting}, such that the user obtains no more information regarding the database other than the requested file. 
In~\cite{chor1995private,chor1998private,gertner1998protecting}, the database is modeled as a bit string, and the user wishes to retrieve a single bit.
In these works, the communication cost is measured as the sum of the transmission at the querying phase from user to servers and at the downloading phase from servers to user.

%

When the file size is significantly large and the target is to minimize the communication cost of only the downloading phase, the metric of the downloading cost is defined as the number of bits downloaded per bit of the retrieved file, and the reciprocal of which is named the \emph{PIR capacity}. 
A series of recent works derives information-theoretic limits of various versions of the PIR problem
\cite{sun2017capacity,sun2016colluding,sun2016SPIR,banawan2016capacity,wang2016symmetric,banawan2017multi} {\it etc}. 
The leading work in the area is by Sun and Jafar\cite{sun2017capacity}, where the authors find the capacity of the PIR problem with replicated databases.
In subsequent works by Sun and Jafar~\cite{sun2016colluding,sun2016SPIR}, the PIR capacity with duplicated databases, duplicated databases with colluding servers, SPIR with duplicated (non-colluding) databases are derived.
In~\cite{banawan2016capacity,banawan2017multi}, Banawan and Ulukus find the capacity of the PIR problem with coded databases, and multi-message PIR with replicated databases.
In our previous work~\cite{wang2016symmetric}, we derive the capacity of the SPIR problem with coded databases.


Considering another cost in storage systems, {\it i.e.} storage overhead, a series of works studies schemes and information limits for various PIR problems with coded databases~\cite{shah2014one,fazeli2015pir,chan2015private,tajeddine2016private,banawan2016capacity,wang2016symmetric,freij2016private}.
In~\cite{shah2014one}, PIR is achieved by downloading one extra bit other than the desired file, given that the number of storage nodes grows with file size, which can be impractical in some storage systems.
In~\cite{fazeli2015pir}, storage overhead can be reduced by increasing the number of storage nodes.
In~\cite{chan2015private}, tradeoff between storage cost and downloading cost is analyzed. Subsequently in~\cite{tajeddine2016private}, explicit schemes which match the tradeoff in~\cite{chan2015private} are presented.
It is worth noting that in~\cite{banawan2016capacity}, the capacity of PIR for coded database is settled, which improves the results in~\cite{chan2015private,tajeddine2016private}.
In our previous work~\cite{wang2016symmetric}, the capacity of SPIR for coded database is derived. Recently in~\cite{freij2016private}, the authors present a framework for PIR from coded database with colluding servers, and conjectured the capacity of the problem, which is settled more recently in~\cite{sun2017private}.

In this work, we generalize our previous work~\cite{wang2016symmetric} on SPIR for MDS-coded database to the case with $T$-colluding servers, hence called MDS-TSPIR. We show that the capacity conjectured in~\cite{freij2016private} for PIR from coded database with colluding servers (and settled in~\cite{sun2017private}) is actually the capacity for the linear SPIR version of the problem.
In analogy to previous works on SPIR~\cite{sun2016SPIR,wang2016symmetric}, in the non-trivial context where the database comprises at least two files, storage nodes need to share common randomness which is independent to the database and meanwhile unavailable to the user.
In particular, we derive the capacity of linear SPIR for an $(N,M)$-MDS coded database where any $T$ servers may collude.
We also derive a lower bound on the amount of common randomness needed to assure nonzero capacity. 
We conjecture that our result holds for general MDS-TSPIR schemes, 
which would reduce to our previous result in~\cite{wang2016symmetric} for non-colluding servers with $T=1$, and the result in~\cite{sun2016SPIR} for replicated non-colluding databases with $T=1$ and $M=1$.

\section{Model}
\subsection{Notations}
Let $[m:n]$ denote the set $\{m, m+1, \dots, n\}$ for $m \leq n$. For the sake of brevity, denote the set of random variables $\{ X_m, X_{m+1}, \dots, X_n\}$ by $X_{[m:n]}$ . 
The transpose of matrix $\bfA$ is denoted by $\bfA^{\textrm{T}}$.

\subsection{Problem Description}
\noindent{\bf Database:}
A database comprises $\nFile$ independent files, denoted by $W_1, \dots, W_{\nFile}$. Each file consists of $\sFile$ symbols drawn independently and uniformly from the finite field $\Fq$. Therefore, for any $k \in [1:\nFile]$, 
\begin{equation}
H(W_k)=L \log{q} \quad ; \quad H(W_1, \dots, W_{\nFile}) = KL \log{q}. \nonumber
\end{equation}

\noindent{\bf Storage:} 
The database is stored in a distributed storage system consisting of $\nNode$ servers (nodes) by an $(\nNode, \nSysNode)$-MDS storage code.  The data stored at node-$n$ is denoted by $D_n$. Note that for any $M$ nodes $\{n_1, \dots, n_M\} \subset [1:N]$, the data $\{D_{n_1},\dots, D_{n_M}\}$ are linearly and stochastically independent.
Furthermore, they can exactly recover the whole database, {\it i.e.,}
\begin{equation}
H(D_{n_1},\dots, D_{n_M}) = H(W_1, \dots, W_{\nFile}) = KL \log{q}, \nonumber
\end{equation}
\begin{equation}
H(W_1,\ldots,W_K|D_{n_1},\ldots,D_{n_M}) = 0. \nonumber
\end{equation}

\noindent{\bf User queries:}
A user wants to retrieve a file $W_{\findex}$ from the database, where the desired file index $\findex$ is drawn from $[1:\nFile]$. 
The MDS storage code is known to the user. In addition to this, the user has no knowledge of the stored data.
Let $\calU$ denote a random variable privately generated by the user, which represents the randomness of the query scheme followed by the user. The random variable $\calU$ is independent of the database and the desired file index. Based on the realization of the desired file index $k$ and the realization of $\calU$, the user generates and sends queries to all nodes, where the query received by node-$n$ is denoted by $Q_{n}^{[k]}$. Let $\queries = \left[Q_{n}^{[k]}\right]_{n \in [1:\nNode], k \in [1:\nFile]}$ denote the complete query scheme, namely, the collection of all queries under all cases of desired file index.
We have that $H(\queries | \calU) = 0$.

{
\noindent{\bf Node common randomness:}
Let random variable $S$ denote the common randomness shared by all nodes, the realization of which is known to all nodes but unavailable to the user. 
The common randomness is utilized to protect database-privacy~\eqref{eqn:data_privacy} below.
For any node $n \in [1:N]$, a random variable $S_n$ is generated from $S$, which is used in the answer scheme followed by node $n$.
Hence, $H(S_1, \dots, S_n | S)=0$.
}

\noindent{\bf Node answers:}
Based on the received query $Q_n^{[k]}$, the stored data $D_n$, and the random variable $S_n$ generated from the common randomness, each node sends an answer $\answer_{n}^{[k]}$ to the user.
Specifically for the linear schemes discussed in this work, the answers are generated by taking the inner product of the query and stored data, then plus the random variable $S_n$ used by the node, {\it i.e.} $A_n^{[k]} = \langle Q_n^{[k]}, D_n \rangle + S_n$.

\noindent{\bf $T$-private SPIR:}
Based on the received answers $\answer_{[1:\nNode]}^{[k]}$ and the query scheme $\queries$, the user shall be able to decode the requested file $W_{k}$ with zero error. 
Any set of $T$ nodes may collude to guess the requested file index, by communicating their interactions with the user.
Two privacy constraints must be satisfied:
\begin{itemize}
\item \emph{User-privacy:} any $T$ colluding nodes shall not be able to obtain any information regarding the identity of the requested file, {\it i.e.,}
	\begin{equation}
		I(\findex ; Q_{\calT}^{[\findex]}, A_{\calT}^{[\findex]}, D_{\calT}, S ) = 0,  \forall \calT \subset [1:\nNode], |\calT|=T. \label{eqn:user_privacy}
	\end{equation}
\item \emph{Database-privacy:} the user shall learn no information regarding other files in the database, that is, defining $W_{\bar{\findex}}= \{ W_1, \dots, W_{\findex-1}, W_{\findex+1}, \dots, W_{\nFile}\}$, 
	\begin{equation}
		I(W_{\bar{\findex}} ; \answer_{[1:\nNode]}^{[\findex]}, \queries, \findex) = 0. \label{eqn:data_privacy}
	\end{equation}
\end{itemize}

We call a $T$-private SPIR scheme a \emph{good} scheme if the user can successfully decode the desired file, with both user-privacy and database-privacy constraints guaranteed.
We use the same definition as in~\cite{wang2016symmetric} for rate and capacity of $T$-private SPIR schemes.
\begin{definition} 
The rate of a $T$-private SPIR scheme is the number of information bits of the requested file retrieved per downloaded answer bit. By symmetry among all files, for any $k \in [1:K]$,
\begin{equation}
R_{T\textrm{-SPIR}}^{(N,M)\textrm{-MDS}} \triangleq \frac{H(W_{k})}{\sum_{n=1}^{\nNode} H(A_n^{[k]})}. \nonumber
\end{equation}
The capacity $C_{T\textrm{-SPIR}}^{(N,M)\textrm{-MDS}}$ is the supremum of $R_{T\textrm{-SPIR}}^{(N,M)\textrm{-MDS}} $ over all $T$-private SPIR schemes for $(N,M)$-MDS storage codes.
\end{definition}

\begin{definition} 
The secrecy rate is the amount of common randomness shared by the storage nodes relative to the file size, by symmetry among all files,
\begin{equation}
\rho_{T\textrm{-SPIR}}^{(N,M)\textrm{-MDS}} \triangleq \frac{H(S)}{H(W_{k})}. \nonumber
\end{equation}
\end{definition}

\section{Main Result} \label{sec:main}
When there is only one file in the database, the two privacy constraints become trivial. When $K \geq 2$, $T$-private SPIR is non-trivial and our main result is summarized below.

\begin{theorem}
For symmetric private information retrieval from a database with $K \geq 2$ files which are stored at $N$ nodes with an $(N,M)$-MDS storage code, where any $T$ nodes may collude, the capacity of linear schemes is 
\begin{equation}
C_{T\textrm{-SPIR}}^{(N,M)\textrm{-MDS}} = 
\begin{cases}
1-\frac{M+T-1}{N}, & \text{if } \rho_{T\textrm{-SPIR}}^{(N,M)\textrm{-MDS}} \geq \frac{M+T-1}{N-M-T+1}\\
0, & \text{otherwise}
\end{cases}
. \nonumber
\end{equation} 
\label{thm:main}
\end{theorem} 

\begin{remark}
\begin{enumerate}
\item The capacity derived holds for linear schemes. For general schemes, the rate achieved by the linear scheme in Section~\ref{sec:achieve} provides an upper bound on the achievable rate. In the proof of the converse part in Section~\ref{sec:converse}, only Lemma~\ref{thm:Lemma1} relies on the constraint that the scheme is linear in the form of $A_n^{[k]} = \langle Q_n^{[k]}, D_n \rangle + S_n$. We conjecture that the Lemma~\ref{thm:Lemma1} and hence the capacity also hold for general schemes.

\item When $T=1$, that is, the storage nodes do not collude, the result of Theorem~\ref{thm:main} reduces to the capacity of MDS-SPIR (for general schemes) derived in our previous work~\cite{wang2016symmetric}.

\item When $M=1$, that is, the database is replicated at each server, the result reduces to $1-\frac{T}{N}$. This is the capacity of general TSPIR, which can be easily proved because the version of Lemma~\ref{thm:Lemma1} for $M=1$ follows directly from user-privacy for general schemes. It is proved in~\cite{sun2016colluding} that the capacity of TPIR is $\frac{1- \frac{T}{N}}{1 -  (\frac{T}{N})^K}$. When $K \to \infty$, the capacity of TPIR approaches the capacity of TSPIR.

\item The problem of MDS-TPIR with no requirement on the database-privacy is directly related to the SPIR version studied in this work. In the first version of~\cite{freij2016private}, the formula $\frac{1-\frac{M+T-1}{N}}{1-(\frac{M+T-1}{N})^K}$ is conjectured to be the capacity of MDS-TPIR, which is disproven in~\cite{sun2017private} by several counterexamples. In Theorem 5 in~\cite{sun2017private}, it is shown that if $N < M+T$, as $K \to \infty$ the capacity of MDS-TPIR decays and converges to $0$. This matches with our result that to achieve positive rate for MDS-TSPIR, the number of servers $N$ should be at least $M+T$. (As with previous works for various scenarios, the PIR capacity reduces to the SPIR capacity by letting $K \to \infty$.) However, the capacity of MDS-TPIR and the capacity of general MDS-TSPIR remain open.
\end{enumerate}
\end{remark}

\section{Converse} \label{sec:converse}
In this section, we show the converse part of Theorem~\ref{thm:main}. 
For any set of nodes, it is direct that given the queries they receive, their answers are independent of the other queries (to other nodes and/or with other desired file indexes), {\it i.e.} $H(A_{\nodeset}^{[k]} | \queries,  Q_{\nodeset}^{[k]}) = H(A_{\nodeset}^{[k]} |  Q_{\nodeset}^{[k]})$ for any $\nodeset \subset [1:\nNode]$. Lemma~\ref{thm:Lemma2} below states that the same holds conditioned on the requested file $W_k$.

\begin{lemma} \label{thm:Lemma2}
For any set of nodes $\nodeset \subset [1:\nNode]$, 
	\begin{equation}
		H(A_{\nodeset}^{[k]} | \queries, W_k, Q_{\nodeset}^{[k]}) = H(A_{\nodeset}^{[k]} | W_k, Q_{\nodeset}^{[k]}). \nonumber
	\end{equation}	
\end{lemma}
\noindent {\it Proof:}
We first show that $I(A_{\nodeset}^{[k]}; \queries | W_k, Q_{\nodeset}^{[k]}) \leq  0$, as follows
\begin{align*}
 I(A_{\nodeset}^{[k]}; \queries | W_k, Q_{\nodeset}^{[k]}) 
& \leq I(A_{\nodeset}^{[k]}, W_{[1:K]}, \secrecy ; \queries | W_k, Q_{\nodeset}^{[k]}) \\
& \stackrel{(a)}{=} I(W_{[1:K]}, \secrecy ; \queries | W_k, Q_{\nodeset}^{[k]}) \\
& \leq I(W_{[1:K]}, \secrecy ; \queries) = 0,
\end{align*}
where $(a)$ holds because the answers are deterministic functions of the database, common randomness, and the queries.
In the last step, $I(W_{[1:K]}, \secrecy ; \queries) = 0$ holds because the queries do not depend on the database and  common randomness.

On the other hand, it is immediate that $I(A_{\nodeset}^{[k]}; \queries | W_k, Q_{\nodeset}^{[k]}) \geq  0$. Therefore, $H(A_{\nodeset}^{[k]}| W_k, Q_{\nodeset}^{[k]}) =  H(A_{\nodeset}^{[k]} | \queries, W_k, Q_{\nodeset}^{[k]})$.
\hfill $\Box$
%

{
In Lemma~\ref{thm:noinfoS} below, we argue that for optimal schemes, the user shall not be able to obtain any information about the shared common randomness $S$ from the received answers.

\begin{lemma} \label{thm:noinfoS}
If there is a good scheme where the user can obtain some information about $S$, {\it i.e.} $H(S | A_{[1:N]}^{[k]}, \queries) < H(S|\queries)$, the scheme can be modified to a good scheme such that the user downloads less (or the same amount of) information bits, where the user learns no information about $S$.
\end{lemma}
\noindent {\it Proof:}
Without loss of generality, assume that $S$ is uniformly distributed, that is, the nodes use the lease possible amount of information bits to generate the shared common randomness.
If $H(S | A_{[1:N]}^{[k]}, \queries) < H(S|\queries) = H(S)$, or equivalently $I(S;  A_{[1:N]}^{[k]}| \queries)) >0$, the conditional distribution of $S$ given the answers is no longer uniformly distributed.
Because from the answers, the user can decode $W_k$, and the user knows the queries, the uncertainty of the answers only lies in the symbols of the other files $W_{\bar{k}}$ and the common randomness $S$. The purpose of $S$ is to randomize the distribution of $W_{\bar{k}}$, such that the distribution of $W_{\bar{k}}$ is still uniformly distributed to the user (by database-privacy).
If the distribution of $S$ conditioned on the queries and answers is not uniformly distributed, the scheme can be modified by reducing the amount of information bits for the common randomness, denote $\tilde{S} = g(S)$ where $g(S)$ is some function of the $S$ used in the original scheme, such that $H(\tilde{S}) = H(S | A_{[1:N]}^{[k]}, \queries)< H(S)$. For the modified scheme, the nodes use the same functions as before (in the original scheme $A_n^{[k]} = f(Q_n^{[k]}, D_n, S)$) but with $\tilde{S}$ to generate the answers, that is $\tilde{A}_n^{[k]} = f( Q_n^{[k]}, D_n, \tilde{S}) =  f( Q_n^{[k]}, D_n, g(S))$.
It can be checked that the scheme is still a good scheme, with less information bits as shared common randomness and less (or at most the same) information bits downloaded. 
The user can still decode $W_k$ successfully, because $\tilde{A}_{[1:N]}^{[k]}$ are generated from less randomness than ${A}_{[1:N]}^{[k]}$ by replacing $S$ with $\tilde{S}$, hence $H(W_k | \tilde{A}_{[1:N]}^{[k]}, \queries) \leq H(W_k| {A}_{[1:N]}^{[k]}, \queries) = 0$. User-privacy is guaranteed by the design of the query scheme, which is not changed. 
To see that database-privacy is still preserved, in the original scheme the user learns some function of $S$, say $l(S)$. Knowing $l(S)$, the database-privacy is guaranteed for the original scheme, hence $I(W_{\bar{k}} ; A_{[1:N]}^{k} | l(S) , \queries) =   0$. For any realization of $l(S)$, the user learns no information of $W_{\bar{k}}$ from the answers. Hence, in the modified scheme where $l(S)$ is set to be some deterministic value, the user still learns no information about $W_{\bar{k}}$.

For linear schemes that are the focus of in this work, it is easy to see that if the user can solve some part of $S$ from the received linear combinations, these solved common randomness are useless in protecting the database $W_{\bar{k}}$, hence are redundant information downloaded. For general functions defined on the finite field, by Lemma II.4. in~\cite{huang2012polynomials}, the function is equivalent to some restriction of a polynomial function. With similar arguments, if the user can solve or obtain any information about $S$, the answers can be modified by using the same functions but less information bits of $S$.

\hfill $\Box$

\begin{remark}
Allowing the user to be able to learn some information about $S$ does not conflict the privacy constraints. For example, the user can send null queries, and receive the $S_n$'s as answers, hence learn some information about the shared common randomness $S$. However, these $S_n$'s received by the user cannot be used by the nodes to protect the database-privacy in transmissions where the database content is involved. Furthermore, the information downloaded by the user about $S_n$'s are redundant in the sense of optimal downloading cost. In Lemma~\ref{thm:noinfoS} above, we formally argue that for optimal schemes, we can require the user to learn no information about $S$, because for the converse, we look for the lowest upper bound of the rate.
\end{remark}

}

%
%
%
%
%
%
%
%
%

{
In Lemma~\ref{thm:Lemma1} below, we show that for any optimal good linear scheme, the entropy of the answers conditioned on the queries from any set of $M+T-1$ nodes are symmetric among all file indexes, and the same holds if also conditioned on the requested file.
}

\begin{lemma} \label{thm:Lemma1}
For any optimal good linear scheme, and for any set of nodes $\nodeset \subset [1:\nNode]$ with size $|\nodeset| = M+T-1$,
	\begin{equation} 
		H(A_{\nodeset}^{[k]} |  Q_{\nodeset}^{[k]}) = H(A_{\nodeset}^{[k']}|  Q_{\nodeset}^{[k']}), \label{eqn:Lemma1_2}
	\end{equation}
	\begin{equation} 
		H(A_{\nodeset}^{[k]} | W_k, Q_{\nodeset}^{[k]}) = H(A_{\nodeset}^{[k']} | W_k, Q_{\nodeset}^{[k']}). \label{eqn:Lemma1_1}
	\end{equation}	
\end{lemma}
\noindent{\it Proof:}
Without loss of generality, assume $\nodeset = [1:M \! + \! T \! - \! 1]$.

{\it Proof of~\eqref{eqn:Lemma1_2}:}
From user-privacy~\eqref{eqn:user_privacy}, for any set $\calT \subset [1:N]$ with size $|\calT| \leq T$, $I(\findex; A_{\calT}^{[\findex]}, Q_{\calT}^{[\findex]}) = 0$, hence $H(A_{\calT}^{[k]}, Q_{\calT}^{[k]})=H(A_{\calT}^{[k']}, Q_{\calT}^{[k']})$. Similarly, $I(\findex;  Q_{\calT}^{[\findex]}) = 0$, therefore $H( Q_{\calT}^{[k]})=H(Q_{\calT}^{[k']})$.
From the above,  we have that $H(A_{\calT}^{[k]}| Q_{\calT}^{[k]})=H(A_{\calT}^{[k']}| Q_{\calT}^{[k']})$.

For an $(\nNode, \nSysNode)$-MDS storage code, the data stored at any set of $M$ nodes are linearly independent. Furthermore, because the files in the database are statistically independent, the data stored at any $M$ nodes are also statistically independent. (See Lemma 1 in~\cite{sun2016colluding} and Lemma 2 in~\cite{banawan2016capacity} for a proof.)
For any node $n$, the answer $A_n^{[k]}$ is a deterministic function of the query $Q_n^{[k]}$, the stored data $D_n$, and the random variable $S_n$ generated from node common randomness. 
We first argue that for optimal schemes, given the queries $Q_{[1:M]}^{[k]}$, the answers $A_{[1:M]}^{[k]}$ are statistically independent.
Firstly, $H(A_{[1:M]}^{[k]} | Q_{[1:M]}^{[k]}) = H(A_{[2:M]}^{[k]} | Q_{[1:M]}^{[k]}) + H(A_{1}^{[k]} | Q_{[1:M]}^{[k]}, A_{[2:M]}^{[k]})$.
Given the queries $Q_{[2:M]}^{[k]}$, the answers of nodes $[2:M]$ depend only on the data $D_{[2:M]}$ and the common randomness $S$, which are independent with $Q_1^{[k]}$. Hence $H(A_{[2:M]}^{[k]} | Q_{[1:M]}^{[k]}) = H(A_{[2:M]}^{[k]} | Q_{[2:M]}^{[k]})$.
The answer of each node is a deterministic function of the query received, the data stored at the node, and the common randomness, hence we write $A_1^{[k]}$ as $A_1^{[k]}(Q_1^{[k]}, D_1, S)$. Given $Q_1^{[k]}$, $A_1^{[k]}$ depends on $D_1$ and $S$, where $D_1$ is independent with $(Q_{[2:M]}^{[k]}, A_{[2:M]}^{[k]})$ by the MDS storage.
If $H(A_{1}^{[k]} | Q_{[1:M]}^{[k]}, A_{[2:M]}^{[k]})$ is strictly smaller than $H(A_{1}^{[k]} | Q_{1}^{[k]}) $, the only possibility is that $(Q_{[2:M]}^{[k]}, A_{[2:M]}^{[k]})$ infer some information about $S$. Hence, the user can download less information from node $1$, because the user shares some information about $S$ with node $1$ now. This contradicts the assumption that the scheme is optimal.
Therefore, $H(A_{1}^{[k]} | Q_{[1:M]}^{[k]}, A_{[2:M]}^{[k]}) = H(A_{1}^{[k]} | Q_{1}^{[k]}) $ and $H(A_{[1:M]}^{[k]} | Q_{[1:M]}^{[k]}) = H(A_{[2:M]}^{[k]} | Q_{[2:M]}^{[k]}) + H(A_{1}^{[k]} | Q_{1}^{[k]})$. Repeat the same steps proves $H(A_{[1:M]}^{k} |  Q_{[1:M]}^{k}) = \sum_{i=1}^{M} H(A_{i}^{[k]}|  Q_{i}^{[k]})$. This holds for any set with at most $M$ nodes.


Fix a set of $T$ nodes, we can neglect the file index in the queries and answers because of user-privacy. 
We argue that for optimal linear schemes, the following equation holds,\footnote{We conjecture that this holds also for general MDS-TSPIR schemes.}

\begin{equation}
I(A_{[1:M-1]}^{[k]}; A_{[M:M+T-1]}|Q_{[1:M-1]}^{[k]}, \! Q_{[M:M+T-1]}) \! = \! 0. \label{eq:withoutwk}
\end{equation}

{
Recall that for linear schemes, for any $n \in [1:M-1]$, $A_n^{[k]} = \langle Q_n^{[k]}, D_n \rangle + S_n$, and for any $n \in [M:M\! + \! T \! - \! 1]$, $A_{n} =  \langle Q_{n}, D_{n} \rangle + S_{n}$. Considering $D_{[1:M]}$ as the basis of the storage, which spans the storage of all nodes, note that for any $n \in [M+1,M+T-1]$, $D_{n}$ is linearly and statistically independent with $D_{[1:M-1]}$, because it contains one independent dimension $D_M$. The argument below is based on this independence, and also the independence of the symbols in $D_{[1:M]}$.
Specifically, for any $n \in [M+1,M+T-1]$, $D_n = \sum_{i=1}^{M} \delta_n^i D_i$ for some coefficients $\delta_n^i$'s.
If equation~\eqref{eq:withoutwk} does not hold, it means that the symbols of $D_M$ can be ``cancelled" from $A_{[M:M+T-1]}$. Hence, the rank of $Q_{[M:M+T-1]}$ is at most $T-1$. In other words, there exists $a_M, \dots, a_{M+T-1}$ which are not all zeros, such that $\sum_{n=M}^{M+T-1} a_n \delta_n^M Q_n = \vec{0}$ (where $\delta_M^M =1$).
Furthermore, note that $D_{[1:M-1]}$ are all composed of independent symbols, in both the dimension along the storage nodes $[1:M-1]$ and the dimension of the symbols stored at each node. If equation~\eqref{eq:withoutwk} does not hold, after forming a sum $\sum_{\textrm{for some } m \in [1:M-1]} \bar{Q}_m D_m + \bar{S}_1$ from $A_{[M:M+T-1]}$, where $\bar{Q}_m = \sum_{n=M}^{M+T-1} a_n \delta_n^m Q_n$, the vector $\bar{Q}_m$, if it does not equal a zero vector, should be aligned with the vector $Q_m^{[k]}$.
Let the sum formed from $A_{1:M-1}^{[k]}$ equals $\sum_{\textrm{for some } m \in [1:M-1]} \bar{Q}_m D_m + \bar{S}_2$, the user downloaded redundant information, regardless of whether $\bar{S}_1 = \bar{S}_2$. 
Because if $\bar{S}_1 = \bar{S}_2$, the user downloaded at least one redundant answer from some node in $[1:M-1]$, which he could deduce $\sum_{\textrm{for some } m \in [1:M-1]} \bar{Q}_m D_m + \bar{S}_2$ from the $T$ answers $A_{[M:M+T-1]}$. 
If $\bar{S}_1 \neq  \bar{S}_2$, the user learns some extra information $\bar{S}_1 - \bar{S}_2$ regarding the shared common randomness, which are not useful to the user, hence is also redundant information downloaded.
Hence, if~\eqref{eq:withoutwk} does not hold, the user has downloaded redundant information, which is a contradiction with the assumption that the scheme is optimal.\footnote{The argument is written in details for each answer being one linear function of query, stored data and common randomness. It is direct that if the answer from each node is multiple such linear functions,~\eqref{eq:withoutwk} still holds from the linear and statistical independence of the data-storage.}

}

{
For example, let $M=2$ and $T=2$, w.l.o.g. we choose the node set $\{1,2,3\}$, with storage $\{ D_1, D_2, D_3 = D_1+D_2 \}$.
Given the queries and answers of the three nodes, we can ignore the file index of node 2 and node 3, which means that $Q_2, Q_3, A_2, A_3$ can be the queries and answers with any desired file index. 
If $Q_1^{[k]} = Q_2 = Q_3$ and $S_1+S_2 = S_3$, indeed $I(A_1^{[k]}; A_2, A_3 |Q_1^{[k]}, Q_2, Q_3) >0$ and it does not violate the two privacy constraints. However, the user could download from only two of the three node, and still obtains the same amount of information without violating privacy constraints. 





}

To conclude,
\begin{align*}
& \quad H(A_{[1:M+T-1]}^{[k]} | Q_{[1:M+T-1]}^{[k]}) \\
& = H(A_{[1:M-1]}^{[k]} | Q_{[1:M-1]}^{[k]}, Q_{[M:M+T-1]}, A_{[M:M+T-1]}) \\
& \qquad + H(A_{[M:M+T-1]} | Q_{[1:M-1]}^{[k]}, Q_{[M:M+T-1]}) \\
& = H(A_{[1:M-1]}^{[k]} | Q_{[1:M-1]}^{[k]})  + H(A_{[M:M+T-1]} | Q_{[M:M+T-1]}) \\
& = \sum_{n=1}^{M-1} H(A_n^{[k]} | Q_n^{[k]})+ H(A_{[M+1:M+T-1]} | Q_{[M+1:M+T-1]}) \\
& = \sum_{n=1}^{M-1} H(A_n^{[k']} | Q_n^{[k']})+ H(A_{[M+1:M+T-1]}^{[k']} | Q_{[M+1:M+T-1]}^{[k']}) \\
& = H(A_{[1:M+T-1]}^{[k']} | Q_{[1:M+T-1]}^{[k']}) 
\end{align*}

{\it Proof of~\eqref{eqn:Lemma1_1}:} Similarly as in the proof of~\eqref{eqn:Lemma1_2}, for any set $\calT$ with no more than $T$ nodes, by user-privacy $H(A_{\calT}^{[k]}| W_k, Q_{\calT}^{[k]})=H(A_{\calT}^{[k']}| W_k, Q_{\calT}^{[k']})$. 
Because any $T$ nodes should not be able to distinguish the requested file index by conditioning on any part of the database, such as the part of data they store regarding $W_k$.

{
Conditioning on $W_k$, the data stored at any $M$ nodes are still statistically independent. Besides, the symbols stored at each node are also statistically independent, because the files are independent and constructed of independent symbols. For the query vector $Q_n$ and stored data vector $D_n$, let $Q_n (\bar{k})$ and $D_n (\bar{k})$ denote the vectors by setting the entries corresponding to file $W_k$ to zeros.
For example, if there are two files in the database and denote $D_1 = (W_{11}, W_{21})$, correspondingly $Q_1^{[k]} = (Q_1^{[k]}(1), Q_1^{[k]}(2))$, where $Q_1^{[k]}(i)$ is to take the inner product of the part of $W_{11}$. Then $Q_1^{[k]}(\bar{1}) = (\vec{0}, Q_1^{[k]}(2))$.
Let $A_n (\bar{k})$ denote the answer $A_n$ conditioned on $W_k$.

For linear schemes such that $A_n^{[k]} = \langle Q_n^{[k]},D_n \rangle +S_n$, the data $D_{[M+1:M+T-1]}$ are linear combinations of $D_{[1:M]}$. In other words, $D_{[M:M+T-1]}$ all have one dimension aligned with $D_{M}$ which is independent with $D_{[1:M-1]}$.
  Denote the node set $[1:M-1]$ by $\calM'$ and the node set $[M:M+T-1]$ by $\calT$, we argue in the following that the version of equation~\eqref{eq:withoutwk} conditioned also on $W_k$ holds. 
  
 \begin{equation}
I(A_{\calM'}^{[k]}; A_{\calT}|Q_{\calM'}^{[k]}, \! Q_{\calT}, W_k) \! = \! 0. \label{eq:withwk}
\end{equation}
 
Because $W_k$ is conditioned in~\eqref{eq:withwk}, the arguments below are similar as in the proof of~\eqref{eqn:Lemma1_2}, but for the vectors obtained by setting the entries corresponding to $W_k$ to zeros, such as $Q_n (\bar{k})$ and $D_n (\bar{k})$, and the answers conditioned on $W_k$, such as $A_n (\bar{k})$.
  If $I(A_{\calM'}^{[k]} ; A_{\calT} | Q_{\calM'}^{[k]} , Q_{\calT} , W_k) > 0$, $Q_{\calT}(\bar{k})$ have rank smaller than or equal to $T-1$, that is, the data aligned with $D_M (\bar{k})$ can be cancelled among $A_{\calT}$. Otherwise, the answers $A_{\calT}$ are formed of linear combinations of symbols from $D_M (\bar{k})$, which are independent of $D_{[1:M-1]} (\bar{k})$.
  Moreover, because $D_{[1:M-1]} (\bar{k})$ are composed of independent symbols, $I(A_{\calM'}^{[k]} ; A_{\calT} | Q_{\calM'}^{[k]} , Q_{\calT} , W_k) > 0$ infers that there exists some linear combination of $A_{\calT} (\bar{k})$ which equals some linear combination of $A_{\calM'}^{[k]} (\bar{k})$. (Or they differ by some linear combination of $S_n$'s, which can be either a constant or some redundant information to the user.)
Let $LC(\cdot)$ denote a linear combination of the input, if~\eqref{eq:withwk} does not hold, the user can calculate $LC_1 (A_{\calM'}^{[k]}) + LC_2 (W_k)$ from $A_{\calT}$, which are independent of the requested file index. The user can download the same answers $A_{\calT}$ from the node set $\calT$ but actually desire a different file $W_{k'}$. After downloading $A_{\calM'}^{[k']}$ from the node set $\calM'$, the user can calculate the same linear combination $LC_1 (A_{\calM'}^{[k']})$ from the answers, such that the common randomness $S_{\calM'}$ can be cancelled by taking $LC_1 (A_{\calM'}^{[k]}) + LC_2 (W_k) - LC_1 (A_{\calM'}^{[k']})$. (Note that $LC_1 (A_{\calM'}^{[k]}) + LC_2 (W_k)$ are computed from $A_{\calT}$.) By database-privacy, the formula $LC_1 (A_{\calM'}^{[k]}) + LC_2 (W_k) - LC_1 (A_{\calM'}^{[k']})$ which the user can compute should only contain information about $W_{k'}$. Hence, $LC_2 (W_k)$ cannot contain information about $W_{kM}$, the part of $W_k$ stored at node $M$. Therefore, the dimension along with $D_M$ should be ``cancelled" by taking linear combinations of $A_{\calT}$, as in the proof of~\eqref{eqn:Lemma1_2}.
If $LC_1 (A_{\calM'}^{[k]}) + LC_2 (W_k) - LC_1 (A_{\calM'}^{[k']})$ equals zero, the user has downloaded redundant information from $\calM'$, hence the scheme is not optimal.
If $LC_1 (A_{\calM'}^{[k]}) + LC_2 (W_k) - LC_1 (A_{\calM'}^{[k']})$ does not equal zero, because $D_{[1:M-1]}$ are independent, there is at least one node $n \in [1:M-1]$ such that $Q_n^{[k]} - Q_n^{[k']}$ has nonzero entries only corresponding to $W_k$ and $W_{k'}$. Replacing this node with node $M$, the new set of $T$ nodes $\calT_1$ may also collude. The nodes can infer the requested file index by checking whether $Q_{\calT_1} (\bar{k})$ and $Q_{\calT_1} (\bar{k'})$ are linearly dependent, hence contradicting user-privacy.
Hence, if~\eqref{eq:withwk} does not hold, either the user has downloaded redundant information such that it contradicts optimality, or the scheme contradicts user-privacy.

}

{
For example, let $M=2$, $T=2$ and $K=2$ (with $W_1 = (W_{11}, W_{12})$ and $W_2 = (W_{21}, W_{22})$), w.l.o.g. we look at the node set $\{1,2,3\}$, with storage $\{ D_1= [W_{11}, W_{21}], D_2 = [W_{12}, W_{22}], D_3 = [W_{11}+W_{12}, W_{21}+W_{22} ] \}$.
Suppose user wants $W_1$, 
if $I ( A_1^1 ;   A_2, A_3 | Q_1^1, Q_2, Q_3, W_1 ) > 0$, 
the queries must be of the form $Q_1^1 = (Q_1^1(1), Q(2)), Q_2 = (Q_2(1), Q(2)), Q_3 = (Q_3(1), Q(2))$.
The user can download $A_2, A_3$, with $W_2$ be the desired file. In this case, $A_1^2 - A_2 - A_3$ can only contain information regarding $W_2$. Hence, $Q_3(1) = Q_2(1)=Q_1^2(1)$. To summary, the queries with $W_1$ desired are $Q_1^1 = (Q_1^1(1), Q(2)), Q_2 = (Q(1), Q(2)), Q_3 = (Q(1), Q(2))$, the queries with $W_2$ desired are $Q_1^2 = (Q(1), Q_1^2(2)), Q_2 = (Q(1), Q(2)), Q_3 = (Q(1), Q(2))$. If $Q_1^1(1) = Q(1)$ or $Q_1^2(2)=Q(2)$, the user can download less information, only from two of the three nodes. Otherwise, node $1$ and node $2$ may collude, and it can be checked that there is no symmetric query scheme with these $Q_1$ and $Q_2$ such that $W_1$ and $W_2$ are both likely to be the desired file.



}

Hence, with similar steps as the proof of~\eqref{eqn:Lemma1_2}, we have that
$H(A_{[1:M+T-1]}^{[k]} | W_k, Q_{[1:M+T-1]}^{[k]}) = H(A_{[1:M+T-1]}^{[k']} | W_k, Q_{[1:M+T-1]}^{[k']})$.
\hfill $\Box$

\begin{lemma} \label{thm:eqn35}
For any set of nodes $\nodeset \subset [1:\nNode]$ with size $|\nodeset| = M+T-1$,
	\begin{equation}
		H(A_{\nodeset}^{[k]} | W_k, Q_{\nodeset}^{[k]}) = H(A_{\nodeset}^{[k']} |  Q_{\nodeset}^{[k']}).\nonumber
	\end{equation}	
\end{lemma}
\noindent{\it Proof:}
By database-privacy~\eqref{eqn:data_privacy}, $ I(W_{\compkprime} ; A_{[1:\nNode]}^{[k']}, \queries) = 0$.
For any $k \neq k'$, because $W_k \in W_{\compkprime}$, we have
\begin{align*}
0
& = I(W_{k} ; A_{\nodeset}^{[k']}, Q_{\nodeset}^{[k']}) \\
& = I(W_{k} ; A_{\nodeset}^{[k']}| Q_{\nodeset}^{[k']}) +  I(W_{k} ; Q_{\nodeset}^{[k']}) \\
& \stackrel{(a)}{=}   I(W_{k} ; A_{\nodeset}^{[k']}| Q_{\nodeset}^{[k']}) \\
& = H(A_{\nodeset}^{[k']}| Q_{\nodeset}^{[k']}) - H(A_{\nodeset}^{[k']}| W_k, Q_{\nodeset}^{[k']}) \\
& \stackrel{(b)}{=} H(A_{\nodeset}^{[k']}| Q_{\nodeset}^{[k']}) - H(A_{\nodeset}^{k}| W_k, Q_{\nodeset}^{[k]}) ,
\end{align*}
where equality $(a)$ holds because $W_k$ is independent of the queries, and equality $(b)$ follows by~\eqref{eqn:Lemma1_1} in Lemma~\ref{thm:Lemma1}.
\hfill $\Box$

\begin{theorem} \label{thm:converse}
The rate of any linear $T$-private SPIR scheme for an $(N,M)$-MDS coded database is bounded from above by
	\begin{equation}
		R_{T\textrm{-SPIR}}^{(N,M)\textrm{-MDS}}  \leq 1-\frac{M+T-1}{N}. \nonumber
	\end{equation}
\end{theorem}
\noindent{\it Proof:} For any file $W_k$, $k \in [1:K]$, and any set of nodes $\nodeset \in [1:N]$ with size $M+T-1$,
\begin{align*}
H(W_k) 
& = H(W_k|\queries) \\
& \stackrel{(a)}{=} H(W_k|\queries) - H(W_k|A_{[1:\nNode]}^{[k]}, \queries) \\
& = H(A_{[1:\nNode]}^{[k]} | \queries) - H(A_{[1:\nNode]}^{[k]} | W_k, \queries) \\
& \leq H(A_{[1:\nNode]}^{[k]} | \queries) - H(A_{\nodeset}^{[k]} | W_k, \queries, Q_{\nodeset}^{[k]}) \\
& \stackrel{(b)}{=} H(A_{[1:\nNode]}^{[k]} | \queries) - H(A_{\nodeset}^{[k]} | W_k, Q_{\nodeset}^{[k]}) \\
& \stackrel{(c)}{=} H(A_{[1:\nNode]}^{[k]}| \queries) - H(A_{\nodeset}^{[k']} |  Q_{\nodeset}^{[k']}) \\
& \stackrel{(d)}{=} H(A_{[1:\nNode]}^{[k]} | \queries) - H(A_{\nodeset}^{[k]} |  Q_{\nodeset}^{[k]}) \\
& \leq H(A_{[1:\nNode]}^{[k]} | \queries) - H(A_{\nodeset}^{[k]} | \queries),
\end{align*}
where $(a)$ holds because the user should be able to decode $W_k$ from $A_{[1:\nNode]}^{[k]}$ and the query scheme, hence $H(W_k|A_{[1:\nNode]}^{[k]}, \queries)=0$.
Equalities $(b)$ and $(c)$ follow from Lemma~\ref{thm:Lemma2} and Lemma~\ref{thm:eqn35}.
Equality $(d)$ follows from~\eqref{eqn:Lemma1_2} in Lemma~\ref{thm:Lemma1}.

Averaging over all $\nodeset$ with size $M+T-1$, we have that
\begin{equation}
H(W_k) \leq H(A_{[1:\nNode]}^{[k]} | \queries) - \frac{1}{{N \choose M+T-1}}\sum_{\substack{ \nodeset \in [1:N] \\ |\nodeset|=M+T-1}} H(A_{\nodeset}^{[k]} | \queries). \nonumber
\end{equation}

By Han's inequality~\cite{cover2012elements}, $$\frac{1}{{N \choose M+T-1}}\sum_{\substack{ \nodeset \in [1:N] \\ |\nodeset|=M+T-1}} H(A_{\nodeset}^{[k]} | \queries) \geq \frac{M+T-1}{N} H(A_{[1:\nNode]}^{[k]} | \queries).$$

Therefore, 
$R_{T\textrm{-SPIR}}^{(N,M)\textrm{-MDS}} = \frac{H(W_k)}{\sum_{n=1}^{\nNode} H(A_n^{[k]})} \leq \frac{H(W_k)}{H(A_{[1:\nNode]}^{[k]} | \queries)} \leq 1 - \frac{M+T-1}{N}$.
\hfill $\Box$

\begin{theorem} \label{thm:secrecy}
The secrecy rate of any linear $T$-private SPIR scheme for an $(N,M)$-MDS coded database needs to be at least 
	\begin{equation}
		\rho_{T\textrm{-SPIR}}^{(N,M)\textrm{-MDS}} \geq \frac{M+T-1}{N-M-T+1} . \nonumber
	\end{equation}
\end{theorem}
\noindent{\it Proof:} Let $\nodeset$ be any set of nodes with size $M+T-1$, by database-privacy~\eqref{eqn:data_privacy},
\begin{align*}
0 
& =  I(W_{\compk} ; A_{[1:\nNode]}^{[k]} | \queries) \\
& = H(W_{\compk} | \queries) - H(W_{\compk} | A_{[1:\nNode]}^{[k]}  , \queries) \\
& \stackrel{(a)}{=} H(W_{\compk} | \queries, W_k) - H(W_{\compk} | A_{[1:\nNode]}^{[k]}  , \queries, W_k) \\
& = I(W_{\compk} ; A_{[1:\nNode]}^{[k]}  | \queries, W_k) \\
& \geq I(W_{\compk} ; A_{\nodeset}^{[k]} | \queries, W_k) \\
& \stackrel{(b)}{=} \! \! H(A_{\nodeset}^{[k]} | \queries, \! W_k) \! - \!  H(A_{\nodeset}^{[k]} | \queries, \! W_{[1:\nFile]}) \! + \!  H(A_{\nodeset}^{[k]} | \queries, \! W_{[1:\nFile]}, \! \secrecy) \\
& = H(A_{\nodeset}^{[k]} | \queries, W_k) - I(\secrecy ; A_{\nodeset}^{[k]} | \queries, W_{[1:\nFile]}) \\
& \geq  H(A_{\nodeset}^{[k]} | \queries, W_k, Q_{\nodeset}^{[k]}) - H(\secrecy) \\
& \stackrel{(c)}{=}  H(A_{\nodeset}^{[k]} | Q_{\nodeset}^{[k]}) - H(\secrecy) \\
& \geq H(A_{\nodeset}^{[k]} | \queries) - H(\secrecy),
\end{align*}
where $(a)$ holds because  $W_k$ is independent of other files $W_{\compk}$, and from $A_{[1:\nNode]}^{[k]} $ and $\queries$ the user can decode $W_k$.
Equality $(b)$ holds because the answers $A_{\nodeset}^{[k]}$ are deterministic functions of the queries $\queries$, the database $W_{[1:K]}$, and the common randomness $\secrecy$.
Equalities $(c)$ follows from Lemmas~\ref{thm:Lemma2}-\ref{thm:eqn35}.

Averaging over all $\nodeset$, and from the proof of Theorem~\ref{thm:converse},
\begin{align*}
H(\secrecy)
& \geq \frac{1}{{N \choose M+T-1}}\sum_{\substack{ \nodeset \in [1:N] \\ |\nodeset|=M+T-1}} H(A_{\nodeset}^{[k]} | \queries) \\
& \geq \frac{M+T-1}{N} H(A_{[1:\nNode]}^{[k]} | \queries) \\
& \geq \frac{M+T-1}{N \! - \! M \! - \! T \! + \! 1} H(W_k).
\end{align*}

Hence, $\rho_{T\textrm{-SPIR}}^{(N,M)\textrm{-MDS}} = \frac{H(S)}{H(W_k)} \geq \frac{M+T-1}{N-M-T+1} $.
\hfill $\Box$

\section{Achievability} \label{sec:achieve}
In this section, we present a scheme which achieves the capacity with the lowest secrecy rate in Theorem~\ref{thm:main}. 
The storage code is an $(N,M)$-MDS code. To achieve $T$-privacy, we generate queries by using an $(N,T)$-MDS code. The answers generated in this setting are codewords from the Schur-product of the storage code and query generating code. From~\cite{mirandola2015critical}, the minimum dimension of the Schur product of the two MDS codes is $M+T-1$ (if $M+T-1\leq N$), achieved when the two MDS codes are generalized Reed Solomon codes with the same evaluation-point sequence. Indeed, the scheme in~\cite{freij2016private} for PIR with coded colluding databases uses generalized Reed Solomon codes. Our scheme develops that in~\cite{freij2016private} to guarantee database-privacy.

\noindent{\bf Database:}
W.o.l.g, let $L \! = \!  M(N \! - \! M \! - \! T \! + \! 1)$ be the file length,
and let matrix $\bfW$ below represent all $K$ files,
\begin{equation}
{\scriptsize
\bfW = 
\begin{bmatrix}
w_{1,1}^{[1]} & w_{1,2}^{[1]} & \dots & w_{1,M}^{[1]} \\
\vdots      & \vdots     & \ddots & \vdots          \\
w_{N \! - \! M \! - \! T \! + \! 1,1}^{[1]} & w_{N \! - \! M \! - \! T \! + \! 1,2}^{[1]} & \dots & w_{N \! - \! M \! - \! T \! + \! 1,M}^{[1]} \\
\vdots      & \vdots     & \ddots & \vdots   \\
w_{1,1}^{[K]} & w_{1,2}^{[K]} & \dots & w_{1,M}^{[K]} \\
\vdots      & \vdots     & \ddots & \vdots              \\
w_{N \! - \! M \! - \! T \! + \! 1,1}^{[K]} & w_{N \! - \! M \! - \! T \! + \! 1,2}^{[K]} & \dots & w_{N \! - \! M \! - \! T \! + \! 1,M}^{[K]} 
\end{bmatrix}
}
, \nonumber
\end{equation}
where every $N \! - \! M \! - \! T \! + \! 1$ rows correspond to a file.

\noindent {\bf Storage:}
Let $\Lambda = (\lambda_1,\dots,\lambda_N) \in \Fq^N$ where $\lambda_i \neq \lambda_j$. Let $\Phi = (\phi_1, \dots, \phi_N) \in \Fq^N$ where $\phi_i \neq 0$. The generating matrix of the storage code is
\begin{equation}
{\scriptsize
\bfG_S = 
\begin{bmatrix}
1 &  \dots & 1\\
\lambda_1 & \dots & \lambda_N         \\
\vdots     & \ddots & \vdots   \\
\lambda_1^{M-1} & \dots & \lambda_N^{M-1}   
\end{bmatrix}
\cdot diag(\Phi)
}
. \nonumber
\end{equation}
The data stored at the $N$ nodes are generated by
\begin{equation}
[D_1,\dots,D_N] = \bfD = \bfW \cdot \bfG_S. \nonumber
\end{equation}

\noindent {\bf Queries:} Let $\Psi = (\psi_1,\dots,\psi_N) \in \Fq^N$ where $\psi_i \neq 0$. The generating matrix of queries is
\begin{equation}
{\scriptsize
\bfG_Q = 
\begin{bmatrix}
1 &  \dots & 1\\
\lambda_1 & \dots & \lambda_N         \\
\vdots     & \ddots & \vdots   \\
\lambda_1^{T-1} & \dots & \lambda_N^{T-1}   
\end{bmatrix}
\cdot diag(\Psi)
}
. \nonumber
\end{equation}

The user queries $M$ rounds. For each round $r$, $r \in [1:M]$, the user generates $T$ independent uniformly random vectors  $U_1^{(r)}, \dots, U_{T}^{(r)}$ of length $(N \! - \! M \! - \! T \! + \! 1)K$ over $\Fq$. Denote $[U_1^{(r)}, \dots, U_{T}^{(r)}] = \bfU^{(r)}$, the $N$ random vectors corresponding to each node are generated by
\begin{equation}
[\tilde{U}_1^{(r)},\dots,\tilde{U}_N^{(r)}] = \tilde{\bfU}^{(r)} =  \bfU^{(r)} \cdot \bfG_Q. \nonumber
\end{equation}

\begin{table*}[ht!]
\scriptsize
\centering
\begin{tabular}{|c|c|c|c|c|c|c|c|c|}
\hline
$E_1$  & $E_2 $ & $\dots$ &$E_M $ &$E_{\nSysNode+1} \sim E_{2 \nSysNode }$ & $E_{2 \nSysNode +1}\sim E_{3 \nSysNode}$ &$\dots$& $E_{\alpha \nSysNode + 1} \sim E_{(\alpha+1)\nSysNode}$ & $E_{(\alpha+1)\nSysNode+1} \sim E_{\nNode}$ \\
\hline \hline
$ e_1$ &$0$ & &$e_2$ &  $e_{\beta+1}$ & $e_{M+\beta+1}$ & & $e_{(\alpha-1)M+\beta+1}$ & $0$  \\
$ e_2$ &$ e_1 $ & & $e_3 $&  $e_{\beta+2}$& $e_{M+\beta+2}$ & &$e_{(\alpha-1)M+\beta+2}$ &$0$ \\
$\vdots$ &$\vdots$  & &$\vdots $&$\vdots $& $\vdots$ & &$\vdots $ & $\vdots$ \\
$ e_{\beta-1} $&$e_{\beta-2}$ &  $\dots$ &$ e_{\beta} $&$e_{2\beta-1}$ & $e_{M+2\beta-1}$ & $\dots$& $e_{(\alpha-1)M+2\beta-1} $ &$0$\\
$ e_{\beta}$ &$e_{\beta-1}$ & &$0$ & $e_{2\beta}$ & $e_{M+2\beta}$ & &$e_{(\alpha-1)M+2\beta}$ & $0$\\
$0$ & $e_{\beta}$ & &$0$ & $e_{2\beta+1}$ & $e_{M+2\beta+1}$ & &$e_{(\alpha-1)M+2\beta+1}$  &$0$   \\
$\vdots$ &$\vdots$ &  &$\vdots $ & $\vdots $ & $\vdots$ & &$\vdots $  & $\vdots$\\
$0$& $0$ & &$e_1$ &  $e_{\beta+M}$ & $e_{2M+\beta}$ & &$e_{\alpha M +\beta }$  &$0$ \\
\hline
\end{tabular}
\caption{Matrix $E$ when user wants $W_1$ and $N \! - \! M \! - \! T \! + \! 1 > M$ (Case 2).}
\label{tb:E_case2}
\vspace{-1cm}
\end{table*}


W.o.l.g, assume the desired file is $W_1$. The detailed query scheme is presented in two orthogonal cases as follows.

\begin{itemize}
\item {\bf Case 1 ($N \! - \! M \! - \! T \! + \! 1 \leq M$)}
Let $e_i$ denote the unit vector of length $(N \! - \! M \! - \! T \! + \! 1)K$ with a one at the $i$th entry, and zeros at all other entries.
Let matrix 
\begin{equation}
{\scriptsize
\bfE \! = \! \! \! 
\begin{bmatrix}
e_1 & 0 & \dots & e_2 & 0 & \! \dots \!& 0 \\
e_2 & e_1 & \dots & e_3 & 0 & \! \dots \!& 0 \\
\vdots & \vdots & \ddots & \vdots & \vdots & \! \ddots \!& \vdots \\
e_{N-M-T} & e_{N \! - \! M \! - \! T \! - \! 1} & \dots & e_{N \! - \! M \! - \! T \! + \! 1} &  & \! \dots \! & \\
e_{N \! - \! M \! - \! T \! + \! 1} & e_{N-M-T} & \dots & 0 &  & \! \dots  \! &  \\
0 & e_{N \! - \! M \! - \! T \! + \! 1} & \dots & 0 & & \! \dots \! &  \\
\vdots & \vdots & \ddots & \vdots & \vdots & \! \ddots \! & \vdots \\
0 & 0 & \dots & e_1 & 0 & \! \dots \! & 0
\end{bmatrix}
} \nonumber
\end{equation}
be an $M(N \! - \! M \! - \! T \! + \! 1)K \times N$ matrix, where the unit vectors are designed in a shifted way among the first $M$ nodes, to retrieve symbols of $W_1$. The queries are generated by randomizing $\bfE$ from adding random matrices $\tilde{\bfU}^{(r)}$, {\it i.e.,}
\begin{equation}
{\scriptsize
Q_{[1:N]}^{[1]} = [\tilde{\bfU}^{(1)}, \dots, \tilde{\bfU}^{(M)}]^{\textrm{T}} + \bfE
}. \label{eqn:randomQ}
\end{equation}

\item {\bf Case 2 ($N \! - \! M \! - \! T \! + \! 1 > M$)}
In this case, let $\beta = N \! - \! M \! - \! T \! + \! 1 \Mod{\nSysNode}$, and $N \! - \! M \! - \! T \! + \! 1= \alpha \nSysNode +\beta$.
The matrix $\bfE$ is as shown in Table~\ref{tb:E_case2}, where $E_n$ denotes the $n$th column of $\bfE$. 
The queries are generated by randomizing $\bfE$ in the same way as in~\eqref{eqn:randomQ}.
\vspace{-0.3cm}
\end{itemize}

For each node $n$, the query $Q_n^{[1]}$ is the $n$th column of $Q_{[1:N]}^{[1]}$, where every $(N \! - \! M \! - \! T \! + \! 1)K$ entries form a query vector for each round, denoted by $Q_n^{[1],(r)}$, $r \in [1:M]$.

\noindent{\bf Answers:}
Each node receives $\nSysNode$ query vectors, and for each forms the inner product with the stored data vector, resulting in $\nSysNode$ symbols. 
To ensure database-privacy, all nodes share $M(M \! +\! T \! - \! 1)$ independent uniformly random symbols from $\Fq$, denoted by $[S_j^{(r)}]_{j \in [1:M \! + \! T \! - \! 1], r \in [1:M]}$, which are independent of the database and unavailable to the user. The answer sent back to the user by node $n$ at round $r$ is generated by
\begin{equation}
A_n^{[1],(r)} = \langle Q_n^{[1],(r)},D_n \rangle + \phi_n  \psi_n \sum_{j=1}^{M+T-1} S_j^{(r)} \cdot \lambda_n^{j-1}. \nonumber
\end{equation}

Because we use an MDS code of dimension $T$ in the generation of queries, every $T$ nodes receive statistically uniformly random query vectors. Hence, user-privacy is guaranteed.

To show that the user can decode $W_1$ successfully and database-privacy is guaranteed, we illustrate via transmission at round 1 under Case 1.
Let $E_n^{(1)}$ denote the vector of the first $(N \! - \! M \! - \! T \! + \! 1)K$ entries in $E_n$, the inner products $\langle E_n^{(1)},D_n \rangle$ for all $N$ nodes retrieve $N \! - \! M \! - \! T \! + \! 1$ linear combinations of different symbols from $W_1$.
\begin{align*}
A_n^{[1],(1)} 
& = \langle Q_n^{[1],(1)},D_n \rangle + \phi_n  \psi_n \sum_{j=1}^{M+T-1} S_j^{(1)} \cdot \lambda_n^{j-1} \\
& = \langle \tilde{U}_n^{(1)}+E_n^{(1)},D_n \rangle + \phi_n  \psi_n \sum_{j=1}^{M+T-1} S_j^{(1)} \cdot \lambda_n^{j-1} \\
& = \phi_n  \psi_n \sum_{l=1}^{(N \! - \! M \! - \! T \! + \! 1)K} (\bfU^{(1)}_{l,1} \cdot 1+\dots+\bfU^{(1)}_{l,T}\cdot \lambda_n^{T-1}) \\
& \quad \cdot (\bfW_{l,1} \cdot 1 +\dots+\bfW_{l,M}\cdot \lambda_n^{M-1}) + \langle E_n^{(1)},D_n \rangle \\
& \quad + \phi_n  \psi_n \sum_{j=1}^{M+T-1} S_j^{(1)} \cdot \lambda_n^{j-1} \\
& =  \phi_n  \psi_n \sum_{j=1}^{M+T-1} (X_j^{(1)}+S_j^{(1)}) \cdot \lambda_n^{j-1} + \langle E_n^{(1)},D_n \rangle,
\end{align*}
where denoting the columns of $\bfW$ by $\bfW_1, \dots, \bfW_M$,
\begin{equation}
X_j^{(1)}  = \sum_{{\substack{t+m=j+1 \\   1 \leq t\leq T ,   1 \leq m \leq M}}} \langle \bfU_{t}^{(1)}, \bfW_m
\rangle \nonumber
\end{equation}

Hence, at round 1, the user downloads $N$ symbols, which are independent linear combinations of $M \! + \! T \! - \! 1$ unknowns $(X_j^{(1)}+S_j^{(1)})$'s and $N \! - \! M \! - \! T \! + \! 1$ unknowns $\langle E_n^{(1)},D_n \rangle$. Therefore, the user can solve the linear system and obtain $N \! - \! M \! - \! T \! + \! 1$ linear combinations of different symbols from $W_1$. The data-base privacy is be guaranteed because of the common randomness $S_j^{(1)}$. Over all $M$ rounds, the user obtain $M(N \! - \! M \! - \! T \! + \! 1)$ independent linear combinations of symbols of $W_1$, which are sufficient to decode the desired file $W_1$.
\vspace{-0.3em}


\bibliographystyle{IEEEtran}
\bibliography{IEEEabrv,SPIR}

\end{document}